\documentclass[conference]{IEEEtran}
\IEEEoverridecommandlockouts
\usepackage{cite}
\usepackage{amsmath}
\usepackage{amsfonts}
\usepackage{amssymb}
\usepackage{url}
\usepackage{algorithm}
\usepackage{algorithmic}
\usepackage{graphicx}
\usepackage{textcomp}
\usepackage{xcolor}
\usepackage{tabularx}
\usepackage{tikz}
\usepackage{adjustbox}
\usepackage{subcaption}
\usepackage{multirow}
\usepackage{multicol}
\usetikzlibrary{quantikz2}
\usepackage[hidelinks]{hyperref}
\usepackage[capitalise]{cleveref}
\creflabelformat{equation}{#2#1#3}
\def\BibTeX{{\rm B\kern-.05em{\sc i\kern-.025em b}\kern-.08em
    T\kern-.1667em\lower.7ex\hbox{E}\kern-.125emX}}

\makeatletter
\newcommand{\linebreakand}{%
  \end{@IEEEauthorhalign}
  \hfill\mbox{}\par
  \mbox{}\hfill\begin{@IEEEauthorhalign}
}
\makeatother
    
\begin{document}

\title{Optimizing Variational Quantum Circuits Using Metaheuristic Strategies in Reinforcement Learning}

\author{
\IEEEauthorblockN{Michael Kölle} 
\IEEEauthorblockA{\textit{LMU Munich} \\
Munich, Germany \\
michael.koelle@ifi.lmu.de}
\and
\IEEEauthorblockN{Daniel Seidl} 
\IEEEauthorblockA{\textit{LMU Munich} \\
Munich, Germany \\
daniel.seidl@campus.lmu.de}
\and
\IEEEauthorblockN{Maximilian Zorn}
\IEEEauthorblockA{\textit{LMU Munich} \\
Munich, Germany \\
maximilian.zorn@ifi.lmu.de}
\linebreakand
\IEEEauthorblockN{Philipp Altmann}
\IEEEauthorblockA{\textit{LMU Munich} \\
Munich, Germany \\
philipp.altmann@ifi.lmu.de}
\and
\IEEEauthorblockN{Jonas Stein}
\IEEEauthorblockA{\textit{LMU Munich} \\
Munich, Germany \\
jonas.stein@ifi.lmu.de}
\and
\IEEEauthorblockN{Thomas Gabor}
\IEEEauthorblockA{\textit{LMU Munich} \\
Munich, Germany \\
thomas.gabor@ifi.lmu.de}
}

\maketitle

\begin{abstract}
Quantum Reinforcement Learning (QRL) offers potential advantages over classical Reinforcement Learning, such as compact state space representation and faster convergence in certain scenarios. However, practical benefits require further validation. QRL faces challenges like flat solution landscapes, where traditional gradient-based methods are inefficient, necessitating the use of gradient-free algorithms. This work explores the integration of metaheuristic algorithms --- Particle Swarm Optimization, Ant Colony Optimization, Tabu Search, Genetic Algorithm, Simulated Annealing, and Harmony Search --- into QRL. These algorithms provide flexibility and efficiency in parameter optimization. Evaluations in $5\times5$ MiniGrid Reinforcement Learning environments show that, all algorithms yield near-optimal results, with Simulated Annealing and Particle Swarm Optimization performing best. In the Cart Pole environment, Simulated Annealing, Genetic Algorithms, and Particle Swarm Optimization achieve optimal results, while the others perform slightly better than random action selection. These findings demonstrate the potential of Particle Swarm Optimization and Simulated Annealing for efficient QRL learning, emphasizing the need for careful algorithm selection and adaptation.
\end{abstract}


\section{Introduction}

Computer science is advancing significantly, particularly in Quantum Computing and Reinforcement Learning (RL). These developments present new research and application opportunities in industries such as automotive, chemistry, and finance \cite{mckinsey2023quantum}. This work addresses the intersection of these two fields, known as QRL, which combines quantum mechanical principles with machine learning strategies, presenting new challenges and opportunities \cite{Cerezo2022}. 


One major issue in Quantum Computing is the flat solution landscape and vanishing gradients characteristic of variational quantum circuits \cite{McClean2018}, posing significant challenges for gradient-based methods used in conventional RL applications \cite{Goodfellow-et-al-2016}. In quantum environments, gradient-based methods can be less effective due to inadequate or misleading gradient information \cite{McClean2018}, impacting traditional optimization techniques in QRL. Therefore, alternative optimization methods need exploration. Studies like Chen et al.'s \cite{Chen_2022} demonstrate the successful use of genetic algorithms for parameter optimization in QRL. This work aims to systematically compare various gradient-free, metaheuristic optimization approaches in QRL, focusing on evaluating their effectiveness in different QRL scenarios and deriving recommendations for using metaheuristics in QRL.



We evaluate the metaheuristics Simulated Annealing (SA), Particle Swarm Optimization (PSO), Ant Colony Optimization (ACO), Tabu Search (TS), and Harmony Search (HS) for optimizing Variational Quantum Circuits (VQC) parameters and compare them to existing benchmarks of Genetic Algorithms (GAs) for QRL. We test these methods in the Cart Pole environment and a Gridworld navigation task, and their performance will be evaluated based on criteria such as performance, convergence speed, and stability to provide a comprehensive comparison. For full reproducibility of our results, the experiment code is available here\footnote{https://github.com/nEWBOBB/metaheuristicQRL}.

The structure of this work is as follows: \cref{metaheuristiken} presents the selected metaheuristic algorithms. The experimental setup is described in  \cref{experimental_setup}. Experimental results are presented and discussed in \cref{results}. The work concludes with a summary, answers to the research questions, and an outlook in \cref{conclusion}. 

\section{Metaheuristic Optimization Algorithms} \label{metaheuristiken}
Metaheuristic optimization algorithms have proven effective for solving complex optimization problems in classical RL and QRL. For instance, genetic algorithms have already been used to optimize hyperparameters of Deep Q-Networks, significantly improving agent performance \cite{such2018deep}. In QRL, Lockwood and Si and Chen \textit{et al.} successfully applied genetic algorithms to optimize parameters of VQCs, leading to more efficient exploration and enhanced learning performance \cite{lockwood2020reinforcement,Chen_2022}.

This Section focuses on metaheuristic optimization algorithms in QRL. It introduces selected metaheuristics, elaborates the algorithm and discusses their application in QRL.

\subsection{Simulated Annealing}
SA is a probabilistic optimization algorithm inspired by metallurgical processes. Introduced in the 1980s by Kirkpatrick, Gelatt, and Vecchi \cite{kirkpatrick}, and independently by Černý \cite{Cerny}, SA simulates the process of gradually cooling a material to achieve thermodynamic equilibrium. The algorithm starts with a random solution and a high temperature, which gradually decreases, reducing the probability of accepting suboptimal solutions and allowing a finer search near a local optimum. SA's effectiveness lies in overcoming local optima, making it suitable for complex, rugged landscapes where traditional descent algorithms often fail \cite{kirkpatrick}. 
In our implementation SA explores the parameter space by sampling different parameter configurations and then uses gradient descent to approach optimal solutions. The temperature mechanism enables it to escape local optima.


\subsection{Particle Swarm Optimization}
PSO is a stochastic, population-based optimization algorithm inspired by the social behavior of bird flocks and fish schools. Introduced in 1995 by Kennedy and Eberhart, PSO simulates social interaction and communication within a group of individuals (particles) to find optimal solutions in a multidimensional search space \cite{kennedy1995}. Each particle is associated with a potential solution, adjusting its position and velocity based on its own and neighbors' experiences.
PSO is effective for various optimization problems, including function optimization and neural network training, due to its simplicity and ability to solve non-linear problems without gradient information \cite{poli2007particle}. We implement PSO by updating the swarmagents' positions and velocity depending on the individuals personal best found rewards and the best found rewards of the swarm. This way they orientate towards the global best solution.


\subsection{Ant Colony Optimization}
ACO is a probabilistic approach based on ant foraging behavior, developed in the early 1990s by Dorigo \cite{Dorigo_Stuetzle_2004}. Ants explore the solution space, leaving pheromone trails that influence the probability of others choosing the same path. This behavior is modeled to solve optimization problems, particularly discrete ones like route planning. ACO is particularly effective for combinatorial optimization problems like the traveling salesman problem and network design, owing to its ability to find good solutions for complex problems where traditional methods struggle \cite{Dorigo_Stuetzle_2004}. 
ACO is implemented by creating a pheromone-matrix the size of the number of agents times itself. Based on the found rewards of the agents/ants, the matrix gets updated and therefore the probabilities for selecting a specific agent change. The selected agent then gets replicated and slightly mutated, to represent a new agent in the next iteration. This simulates the ant moving into the direction of a strong pheromon source, e.g. a successful other ant in regards to obtaining a higher reward.


\subsection{Tabu Search}
TS, developed by Fred Glover in the late 1980s, uses memory structures that record recently explored solutions to avoid cycles and encourage exploration of new areas \cite{glover1986}. The Tabu list temporarily forbids certain moves, increasing the likelihood of escaping local optima.
TS is effective for scheduling, route planning, and network design due to its ability to quickly find good solutions while deeply exploring the search space \cite{Glover1998}.
We implemented a tabu list, which stores a list of solution candidates, that are not accepted. This is based off of their performance, e.g. worse rewards than other candidates. A neighborlist is compared with the tabu list and non-tabu solution candidates will be taken into account. The tabu list then gets updated with the newly selected candidate, so new solutions can be considered more.

\subsection{Harmony Search}
Harmony Search (HS), developed in 2001 by Zong Woo Geem, Joong Hoon Kim, and G. V. Loganathan, is inspired by the musical process of searching for a perfect harmony \cite{geem}. It uses a harmony memory database to store the best solutions found and generates new solutions by combining random selection, memory consideration, and pitch adjustment.
HS is applied in engineering, transport, and energy distribution for its simplicity, flexibility, and robustness against initial conditions and parameter settings \cite{CEYLAN20082527}.
In our work, HS is implemented by creating a harmony memory, that stores a set amount of best solutions. Based off of those, new solutions are created via a small manipulation, depending on variables like bandwidth. The harmony memory then gets updated, with newly and possibly better found solutions, replacing the current worst solution in the harmony memory in the case of a higher reward being found.


\subsection{Genetic Algorithms}
GAs are evolutionary algorithms simulating natural selection, introduced by John Holland \cite{holland1975adaptation} in the 1970s. They start with a population of potential solutions that evolve toward optimality through selection, mutation, and crossover. Fitter solutions, as determined by a fitness function, are more likely to be chosen for breeding, guiding the population toward better solutions. GAs are effective for complex optimization problems due to their ability to avoid local optima \cite{goldberg1989genetic}.
Chen \textit{et al.} \cite{chen2020variational} integrate GAs with quantum computing to optimize quantum circuits for reinforcement learning tasks, showing GAs' utility in high-dimensional problem-solving. In this work, top candidates are selected for mutation to create new generations.


\section{Experimental Setup} \label{experimental_setup}
This Section provides a overview of the components and methodologies utilized in the study. It includes the specific environments we used for testing, namely the $5\times5$ MiniGrid and Cart Pole environment and the variational quantum circuit architecture. The Section also covers the evaluation metrics and hyperparameter selection.

\subsection{Environments}
We evaluate our selected metaheuristics in two popular RL environments for their effectiveness and adaptability in QRL. We use the $5\times5$ MiniGrid and Cart Pole environment from the OpenAI Gym library \cite{brockman2016openai}.

\subsubsection{$5\times5$ MiniGrid Environment}
The $5\times5$ MiniGrid environment \cite{empty-minigrid} is a simplified, grid-based simulation scenario. We use the \textit{Empty} variant, where an agent navigates a 5x5 grid aiming to reach a target position. The observation space has a dimensionality of 75, derived from $5 \times 5 \times 3$.
In this environment, the Markov Decision Process (MDP) is defined as follows: The state set $\mathcal{S}$ includes all possible agent positions in the 5x5 grid. The action set $\mathcal{A}$ comprises six actions: turn left, turn right, move forward, pick up object, drop object, and use object \cite{empty-minigrid}. Transition probabilities $\mathcal{P}$ are deterministic, moving the agent to the neighboring state unless blocked by a wall or obstacle. The reward function $\mathcal{R}$ assigns a positive reward for reaching the target, decreasing with the number of steps taken, while failure to reach the target results in zero reward \cite{Chen_2022}. The discount factor $\gamma$ is typically high, promoting long-term planning \cite{sutton2018reinforcement}. The objective is to find an optimal policy $\pi^*$ that maximizes the expected cumulative reward.

\subsubsection{Cart Pole Environment}
The Cart Pole environment, also known as the Inverted Pendulum, challenges an agent to balance a pole on a cart by moving the cart left or right. This environment is ideal for demonstrating control of continuous systems and balance tasks \cite{brockman2016openai, farama_cart_pole}.
The MDP is defined as follows: The state set $\mathcal{S}$ includes the cart's position and velocity and the pole's angle and angular velocity. The action set $\mathcal{A}$ consists of moving the cart left or right. Transition probabilities $\mathcal{P}$ are determined by differential equations modeling the system's dynamics. The reward function $\mathcal{R}$ gives +1 for each timestep the pole remains upright, ending the episode if the pole deviates too far or the cart moves out of bounds, with a maximum reward of 500. The discount factor $\gamma$ is usually high, encouraging long-term stability.
The agent's objective is to keep the pole upright for as long as possible by controlling the cart. The dynamics of Cart Pole are modeled by differential equations, representing the continuous nature of states and actions. The reward function $\mathcal{R}: \mathcal{S} \times \mathcal{A} \rightarrow \mathbb{R}$ is designed to encourage balancing the pole near vertical and penalizing deviations.

\subsection{Variational Quantum Circuits}
\begin{figure}[htb]
    \centering
    \adjustbox{width=0.75\linewidth}{
    \begin{quantikz}
        \lstick{\ket{0}} & \qw\gategroup[2,steps=3,style={dashed,rounded
        corners,fill=green!20, inner
        xsep=2pt},background,label style={label
        position=below,anchor=north,yshift=-0.2cm}]{{\sc Encoding}}  & \gate[2, style={inner xsep=0.5cm}]{U(x)} & \qw & \qw & \ctrl{1}\gategroup[2,steps=2,style={dashed,rounded
        corners,fill=blue!20, inner
        xsep=2pt},background,label style={label
        position=below,anchor=north,yshift=-0.2cm}]{{\sc
        Variational Part}} & \gate{R(\alpha_1, \beta_1, \gamma_1)} & \meter{} \\
        \lstick{\ket{0}} & \qw & \qw & \qw & \qw & \targ{} & \gate{R(\alpha_2, \beta_2, \gamma_2)} & \meter{}
    \end{quantikz}
    }
\caption[Quantum Circuit for Cart Pole Environment.]
{Quantum Circuit for Cart Pole Environment of Chen \textit{et al.}\cite{Chen_2022}. The $U(x)$ gate represents the encoding, here realized using Amplitude Encoding. The variational part consists of a CNOT gate, followed by parameterized rotational gates with trainable parameters $\alpha_i$, $\beta_i$, and $\gamma_i$. The variational part is repeated four times. The output is a 2-dimensional tuple $[a,b]$, determining the cart movement direction.}
\label{fig:CPcircuit}
\end{figure}
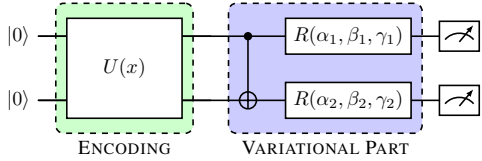

The architecture of a VQC typically consists of three parts: the encoding, the variational part, and the measurement (as seen in \cref{fig:5x5circuit,fig:CPcircuit}). The encoding is responsible to handle classical input and encode it into the Hilbert space. The variational part contains trainable parameters and is optimized by a classical optimizer to fit the target function. In our RL case, the VQC architecture depends on the dimensions of the state and action space of each environment. 
In current NISQ hardware, environment states may be too big to be encoded in the limited number of qubits available. Therefore, the state may be compressed. Low-dimensional environments like Cart Pole do not require dimensionality reduction, and observations can be encoded directly into qubits using, e.g., Amplitude Encoding \cite{Chen_2022}. For higher-dimensional environments (like MiniGrid), Matrix Product States (MPS) can reduce complexity without significant information loss \cite{Chen_2022}. 

\subsubsection{Matrix Product States}
MPS efficiently represent and manipulate quantum states \cite{biamonte2017tensor}. For the $5\times5$ MiniGrid environment, the state vector dimension is 75 ($5 \times 5 \times 3$) \cite{empty-minigrid}. Applying a feature map $\phi(v)$ with features $\in \{1,...,N\}$ reduces this complexity:

\begin{equation}
v \rightarrow |\phi(v)\rangle = |\phi(v_1)\rangle \otimes |\phi(v_2)\rangle \otimes \cdots \otimes |\phi(v_N)\rangle,
\end{equation}

\noindent With each $\phi(v_j)$ as a $d$-dimensional vector. Following Chen et al. \cite{Chen_2022}, we use $d=2$:

\begin{equation}
\phi(v_j) =
\begin{bmatrix}
1 - v_j \
v_j
\end{bmatrix}.
\end{equation}

\noindent MPS transforms the chain of matrices into a state vector, reducing input dimension from 75 to 8 for efficient processing.~\cite{Chen_2022}

\subsubsection{Variational Encoding}
Variational Encoding, combined with MPS, prepares states in quantum circuits. The compressed state vector from MPS serves as input. Hadamard gates create an initial state:

\begin{equation}
\label{eq1}
\ket{\psi_0} = H^{\otimes n} \ket{0}^{\otimes n} = \frac{1}{\sqrt{2^n}} \sum_{i=0}^{2^n-1} \ket{i}.
\end{equation}

\noindent Rotational gates $R_Y$ and $R_Z$ encode the state of the environment:

\begin{equation}
\label{eq2}
Ry(\arctan(x_i)) \otimes Rz(\arctan(x^2_{i})),
\end{equation}

\noindent where $x_i$ is from the compressed state vector. Overall:

\begin{equation}
\label{eq
}
\ket{\psi_{\text{enc}}} = \prod_{i=1}^{n/2} \left(Ry(\arctan(x_i)) \otimes Rz(\arctan(x^2_{i}))\right) \ket{\psi_0}.
\end{equation}

\noindent This encoded state is then used by the VQC for decision-making.

\subsubsection{Amplitude Encoding}
The Cart Pole environment provides a 4-dimensional observation vector representing the cart's position and velocity, and the pole's angle and rotational velocity. Amplitude Encoding translates real vectors into the amplitudes of a quantum state \cite{mottonen2004transformation,schuld2021supervised,larose2020robust}.
An $n$-dimensional real vector $\mathbf{x} = (x_1, \ldots, x_n)$ is mapped to a quantum state $\ket{\psi_\mathbf{x}}$ represented by $\lceil \log_2 n \rceil$ qubits, where the amplitudes are the normalized components of the (possibly zero-padded) input vector:

\begin{equation}
\ket{\psi_\mathbf{x}} = \frac{1}{\sqrt{\sum_{i=1}^n |x_i|^2}} \sum_{i=1}^n x_i \ket{i},
\end{equation}

\noindent For the Cart Pole environment, the 4-dimensional observation requires 2 qubits:

\begin{equation}
\ket{\psi_\mathbf{x}} = \frac{x_1 \ket{00} + x_2 \ket{01} + x_3 \ket{10} + x_4 \ket{11}}{\sqrt{\sum_{i=1}^4 |x_i|^2}}.
\end{equation}


\subsubsection{Variational Quantum Circuit}
VQC is essential in Quantum Machine Learning, especially in QRL. For Cart Pole, Amplitude Encoding is used. For $5\times5$ MiniGrid, MPS reduces state complexity before encoding. VQC uses these states to make decisions, optimizing actions based on encoded states. The action selection process in this work is similar to Quantum Deep Q-Learning.~\cite{Chen_2022}

\begin{figure}[hbt]
    \centering
    \adjustbox{width=\linewidth}{
    \begin{quantikz}
    \lstick{\ket{0}} & \gate{H}\gategroup[8,steps=3,style={dashed,rounded
    corners,fill=green!20, inner
    xsep=2pt},background,label style={label
    position=below,anchor=north,yshift=-0.2cm}]{{\sc
    Encoding}}  & \gate{R_y(\arctan(x_1))} & \gate{R_z(\arctan(x^2_1))} & \ctrl{1}\gategroup[8,steps=9,style={dashed,rounded
    corners,fill=blue!20, inner
    xsep=2pt},background,label style={label
    position=below,anchor=north,yshift=-0.2cm}]{{\sc
    Variational Part}} & \qw & \qw & \qw & \qw & \qw & \qw & \targ{} & \gate{R(\alpha_1, \beta_1, \gamma_1)} & \meter{} \\
    \lstick{\ket{0}} & \gate{H} & \gate{R_y(\arctan(x_2))} & \gate{R_z(\arctan(x^2_2))} & \targ{} & \ctrl{1} & \qw & \qw & \qw & \qw & \qw & \qw & \gate{R(\alpha_2, \beta_2, \gamma_2)} & \meter{} \\
    \lstick{\ket{0}} & \gate{H} & \gate{R_y(\arctan(x_3))} & \gate{R_z(\arctan(x^2_3))} & \qw & \targ{} & \ctrl{1}  & \qw & \qw & \qw & \qw & \qw & \gate{R(\alpha_3, \beta_3, \gamma_3)} & \meter{} \\
    \lstick{\ket{0}} & \gate{H} & \gate{R_y(\arctan(x_4))} & \gate{R_z(\arctan(x^2_4))} & \qw & \qw & \targ{} & \ctrl{1}  & \qw & \qw & \qw & \qw & \gate{R(\alpha_4, \beta_4, \gamma_4)} & \meter{} \\
    \lstick{\ket{0}} & \gate{H} & \gate{R_y(\arctan(x_5))} & \gate{R_z(\arctan(x^2_5))} & \qw & \qw & \qw & \targ{} & \ctrl{1} & \qw & \qw & \qw & \gate{R(\alpha_5, \beta_5, \gamma_5)} & \meter{} \\
    \lstick{\ket{0}} & \gate{H} & \gate{R_y(\arctan(x_6))} & \gate{R_z(\arctan(x^2_6))} & \qw & \qw & \qw & \qw & \targ{} & \ctrl{1} & \qw & \qw & \gate{R(\alpha_6, \beta_6, \gamma_6)} & \meter{} \\
    \lstick{\ket{0}} & \gate{H} & \gate{R_y(\arctan(x_7))} & \gate{R_z(\arctan(x^2_7))} & \qw & \qw & \qw & \qw & \qw & \targ{} & \ctrl{1} & \qw & \gate{R(\alpha_7, \beta_7, \gamma_7)} \\
    \lstick{\ket{0}} & \gate{H} & \gate{R_y(\arctan(x_8))} & \gate{R_z(\arctan(x^2_8))} & \qw & \qw & \qw & \qw & \qw & \qw & \targ{} & \ctrl{-7} & \gate{R(\alpha_8, \beta_8, \gamma_8)}
    \end{quantikz}
    } 
 \caption[Quantum Circuit for $5\times5$ MiniGrid Environment]{Quantum Circuit for $5\times5$ MiniGrid Environment. The $5\times5$ MiniGrid input is reduced to a dimension of 8 using MPS. Eight qubits are initialized and manipulated with Hadamard gates into superposition. The 8-dimensional input is encoded using $R_Y$ and $R_Z$ rotational gates. The variational part involves entangling all qubits and applying rotation gates with trainable parameters $\alpha_i$, $\beta_i$, and $\gamma_i$. Measurements on 6 of 8 qubits represent the action set $\mathcal{A} = 6$, with action probabilities derived using a softmax function (see Chen \textit{et al.} \cite{Chen_2022}).}
\label{fig:5x5circuit}
\end{figure}
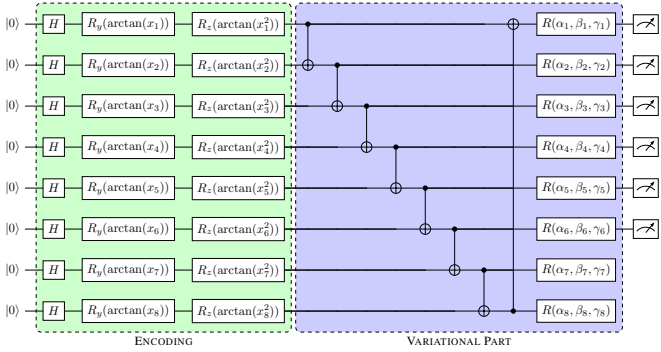

\subsection{Evaluation Metrics}\label{evaluation_criteria}

\subsubsection{Learning Speed} The learning speed $v_l$ indicates how quickly an algorithm can develop effective strategies. It is assessed by the number of episodes ($E$) needed to reach a specific performance level ($P$):

\begin{equation}
v_l = \frac{E}{P}.
\end{equation}

Faster learning speed (i.e. smaller $v_l$) is desirable in resource-limited or real-time conditions.

\subsubsection{Stability} The stability $S$ refers to the consistency of an algorithm's performance across multiple runs, quantified as the standard deviation (i.e. $\sigma$) of performance ($P_i$) over $N$ runs:

\begin{equation}
\sigma = \sqrt{\frac{1}{N} \sum_{i=1}^{N} (P_i - \bar{P})^2}.
\end{equation}

Lower standard deviation (smaller $\sigma$) indicates higher stability.

\subsubsection{Maximum Performance} The maximum performance $P_{max}$ indicates the highest achieved performance (average reward), reflecting the effectiveness of the learning process.

\subsection{Hyperparameter Selection}
We determined the optimal hyperparameters for each metaheuristic algorithm through a grid search, systematically testing different combinations to identify the best configuration for each environment. Each algorithm is allocated the same computation time for fairness.
Agents are initialized with parameter values drawn from a standard normal distribution $\mathcal{N}(0,1)$, scaled by 0.01 for suitable initial values. Each hyperparameter configuration was run with 3 different seeds $\in [0,2]$. A maximum runtime per run of $t_{\text{max}} = 20000$ seconds is set to ensure sufficient iterations and convergence without excessive computation time. The optimal hyperparameters for each metaheuristic can be found in \cref{tab:hyparams}.


\begin{table*}[htbp]
\caption{Optimal Hyperparameter Configurations}
\label{tab:hyparams}
\begin{tabularx}{\textwidth}{|l|X|X|}
\hline
\textbf{Algorithm} & \textbf{$5\times5$ MiniGrid} & \textbf{CartPole} \\
\hline
\textbf{Ant Colony Optimization} & Number of ants = 10, Evaporation rate = 0.9, $\alpha$ = 0.5, $\beta$ = 0.3 & Number of ants = 30, Evaporation rate = 0.95, $\alpha$ = 1.0, $\beta$ = 1.5 \\\hline
\textbf{Harmony Search} & Harmony memory size = 30, Acceptance rate = 0.9, Pitch adjustment rate = 0.3, Bandwidth = 0.1 & Harmony memory size = 100, Acceptance rate = 0.8, Pitch adjustment rate = 0.5, Bandwidth = 0.3 \\\hline
\textbf{Particle Swarm Optimization} & Number of particles = 20, Inertia coefficient = 0.4, Cognitive coefficient = 1.0, Social coefficient = 1.5 & Number of particles = 20, Inertia coefficient = 0.9, Cognitive coefficient = 1.0, Social coefficient = 2.0 \\\hline
\textbf{Simulated Annealing} & Initial temperature = 3000.0, Cooling rate = 0.001 & Initial temperature = 500000.0, Cooling rate = 0.001, Stopping criterion = 0.001 \\\hline
\textbf{Tabu Search} & Tabu list size = 7, Neighborhood size = 20 & Tabu list size = 10, Neighborhood size = 40 \\\hline
\textbf{Genetic Algorithm} & Number of agents = 500, Selected top agents = 10 & Number of agents = 500, Selected top agents = 10 \\\hline
\end{tabularx}
\end{table*}

\section{Results} \label{results}
This section presents the experimental results from applying selected metaheuristic algorithms to QRL problems in the $5\times5$ MiniGrid and Cart Pole environments. The performance of each algorithm is compared to highlight their strengths and weaknesses.

\begin{figure*}[htb]
 \centering
  \subfloat[$5\times5$ MiniGrid]{
   \label{fig:5x5_res}
   \includegraphics[width=0.48\linewidth]{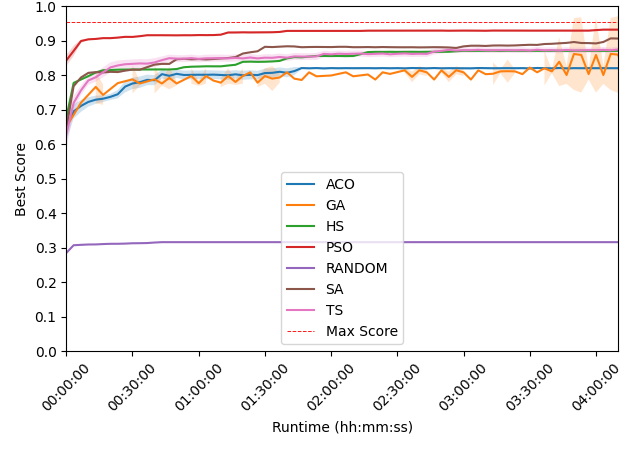}}
  \subfloat[Cart Pole]{
   \label{fig:CP_res}
   \includegraphics[width=0.48\linewidth]{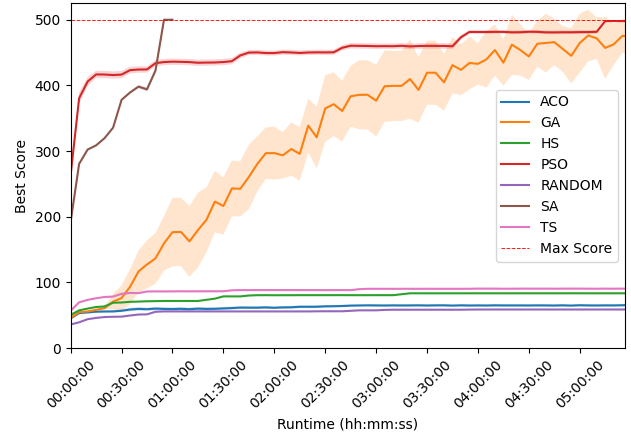}} 
 \caption{Results of each metaheuristic in optimizing the RL agent in the $5\times5$ MiniGrid and Cart Pole environment. To ensure a fair comparison, the number of iterations was adjusted to match an equal runtime. The plots show the highest reward achieved (Y-axis) over wall-clock runtime (X-axis) for each of the 5 runs, with results visualized using the mean and 95\% confidence interval. Each metaheuristic reward was averaged over 5 runs. }
 \label{fig:results}
\end{figure*}

\subsection{Results in the $5\times5$ MiniGrid Environment}
In the $5\times5$ MiniGrid environment, the maximum achievable score is 0.955. \cref{fig:5x5_res} shows the results of each algorithm over a runtime of $T_{max} = 20000$ seconds (approximately 5.5 hours) on a Intel(R) Core(TM) i5-4570 CPU @ 3.20GHz. All algorithms quickly reach average rewards above $0.7$. Except for ACO, all algorithms exceed rewards of 0.8 within 30 minutes. ACO converges to 0.83 after 1.75 hours. HS, TS, and SA behave similarly, reaching between 0.83 and 0.85 after a little over an hour and converging to 0.88 after 2.75 hours. SA finds better solutions earlier, reaching 0.88 after 1.5 hours and converging to 0.9-0.91 after nearly 4 hours. PSO performs best, reaching 0.85 early and 0.9 after less than 15 minutes, converging to 0.93-0.94 in less than 2 hours. At the same time, the GA does not reach near-optimal rewards. As ACO it stays below rewards of 0.8, after roughly 3.5 hours it then finds better solutions, but with a high difference in top rewards over the 5 runs (low stability), ranging from 0.8-0.94. 

These results indicate that PSO excels with quick convergence and high rewards, while ACO is less suitable for this environment, converging slower and achieving lower maximum rewards. All algorithms outperform a random action selection algorithm, as shown in \cref{fig:5x5_res}.


\subsection{Results in the Cart Pole Environment}
In the Cart Pole environment, the maximum score is 500. \cref{fig:CP_res} shows the results of each metaheuristic. ACO performs the worst, converging to a reward of 72 after 2 hours. HS and TS have similar performance, reaching rewards of 70 after 30 minutes and converging to 85 and 90, respectively, after around 3 hours. SA performs well, achieving rewards of 300 within 15 minutes and converging to the maximum reward of 500 in 3 out of 5 runs after 1 hour. PSO performs the best, achieving rewards of 400 within 10 minutes and converging to the maximum reward of 500 after about 5 hours. Other than for the $5\times5$ MiniGrid Environment, the GA starts slow like the bottom 4 algorithms but then quickly finds better solutions over time, getting rewards between 420 and 500 after 5 hours of runtime.

These results show that SA and PSO perform excellently, finding optimal solutions quickly. GA also finds optimal solutions, even if not in all of the runs. HS and TS perform worse, needing longer times and achieving lower rewards. ACO is unsuitable, converging slowly and achieving low rewards.


\subsection{Comparison of Metaheuristic Algorithms}



\begin{table}[htbp]
\centering
\caption{Evaluation criteria results in the $5\times5$ MiniGrid and Cart Pole environments}
\label{tab:combinedresults}
\begin{tabularx}{\linewidth}{|l|XXX|XXX|}
\hline
\multirow{2}{*}{\textbf{Alg.}} & \multicolumn{3}{c|}{\textbf{$5\times5$ MiniGrid}} & \multicolumn{3}{c|}{\textbf{Cart Pole}} \\
\cline{2-7}
 & \textbf{$v_l$} & \textbf{$S$} & \textbf{$P_{max}$} & \textbf{$v_l$} & \textbf{$S$} & \textbf{$P_{max}$} \\
\hline
SA & 2.62 & 0.019 & 0.890 & \textbf{0.0011} & 70.54 & 410.4 \\
PSO & \textbf{1.90} & 0.010 & \textbf{0.931} & 0.0054 & \textbf{4.47} & \textbf{498.0} \\
ACO & 3.19 & \textbf{0.004} & 0.828 & 0.0398 & 5.48 & 65.4 \\
TS & 2.66 & 0.025 & 0.879 & 0.0205 & 10.71 & 90.4 \\
HS & 2.59 & 0.021 & 0.871 & 0.0328 & 2.00 & 83.0 \\
GA & 6.61 & 0.055 & 0,832 & 0.0087 & 29.75 & 475,2 \\
\hline
\end{tabularx}
\end{table}

In the $5\times5$ MiniGrid environment, PSO achieves the highest learning speed, reaching a maximum performance of 0.931 in 6380 seconds (1.77 hours), and shows the highest average performance of 0.931. In the Cart Pole environment, PSO again demonstrates the highest learning speed, achieving a score of 498 in 9760 seconds (2.71 hours), and exhibits the highest average performance of 498. PSO also shows the highest stability in the Cart Pole environment with a standard deviation of 4.47, while ACO has the highest stability in the $5\times5$ MiniGrid environment with a standard deviation of 0.004. The GA exhibits low stability in both environments. PSO's high adaptability is evident as it archives the highest maximum performance and maintains good stability across both environments. SA also displays good adaptability but with higher performance variability in the Cart Pole environment. HS, TS, and ACO perform near-optimal in the $5\times5$ MiniGrid environment but poorly in the Cart Pole environment, indicating lower adaptability.The GA performs less optimal in the $5\times5$ MiniGrid environment but shows high performance in the Cart Pole environment. Overall, PSO demonstrates the highest adaptability and performance in both environments, with the fastest learning speed, highest maximum performance, and good stability, making it particularly suitable for optimizing QRL models across various problems. SA also performs well, especially when high maximum performance is prioritized over consistency.The GA performs well in the Cart Pole environment but shows weaker performance in the $5\times5$ MiniGrid environment. HS, TS, and ACO are less robust, showing acceptable results in specific environments but less adaptability to different problems. It is important to note that optimization with the GA reaches near-optimal solutions in both environments after a prolonged period, i.e. significantly longer than the runtimes shown in this work.

\section{Conclusion} \label{conclusion}
This study evaluated various metaheuristic optimization algorithms for QRL in the $5\times5$ MiniGrid and Cart Pole environments, using criteria such as learning speed, stability, maximum performance, and adaptability. The PSO algorithm demonstrated the highest learning speed, maximum performance, and adaptability in both environments, showing good stability across runs. The SA algorithm also performed well, particularly in terms of maximum performance, though it showed less stability in the Cart Pole environment.The GA demonstrates high performance in the Cart Pole environment but requires longer to reach near-optimal solutions in both environments. Other algorithms, including HS, TS, and ACO, performed acceptably in the $5\times5$ MiniGrid environment but poorly in the Cart Pole environment, indicating lower adaptability.

Future research could explore the performance of metaheuristic algorithms in more complex environments with higher-dimensional state spaces or continuous action spaces. Real-world applications, such as robotics, could also be investigated. Additionally, optimizing algorithm-specific hyperparameters through automated methods could enhance efficiency. Developing adaptive hyperparameter strategies and hybrid approaches that combine different metaheuristic algorithms could further improve performance. Finally, testing these algorithms on actual quantum hardware rather than simulators could uncover new challenges and opportunities in QRL.

\section*{Acknowledgements}
This work is part of the Munich Quantum Valley, which is supported by the Bavarian state government with funds from the Hightech Agenda Bayern Plus.

\bibliographystyle{unsrt}  
\bibliography{main}
\end{document}